\documentclass[fleqn,usenatbib]{mnras}
\usepackage{lastpage}
\usepackage[T1]{fontenc}
\usepackage{lastpage}
\usepackage{graphicx}
\usepackage{verbatim}
\usepackage{txfonts}
\usepackage{float}
\usepackage{color}
\usepackage{amsbsy}
\usepackage{gensymb}
\usepackage{textcomp}
\usepackage{dcolumn}
\usepackage{natbib}
\usepackage{multirow}

\usepackage{xspace}
\usepackage{hyperref}
\usepackage{caption}
\usepackage{orcidlink}
\usepackage{tabularx}

\usepackage{amssymb}	
\usepackage{epsfig}
\usepackage{upgreek}
\usepackage{mathptmx}
\usepackage[utf8]{inputenc}
\usepackage{enumerate}
\usepackage{enumitem}   
\usepackage{longtable,lscape}

\usepackage{grffile}
\usepackage{array,float}
\usepackage{lscape}
\usepackage{pdflscape}
\usepackage{afterpage}
\usepackage{siunitx}
\usepackage{xargs}

\usepackage{setspace}

\usepackage[colorinlistoftodos]{todonotes}
\usepackage{adjustbox}
\usepackage{booktabs}
\usepackage{soul}
\usepackage{ulem}
\usepackage{lscape}

\newcolumntype{d}[1]{D{.}{\cdot}{#1}}
\definecolor{mygray}{gray}{0.6}
\newcolumntype{.}{D{.}{.}{-1}}

\bibpunct[; ]{(}{)}{,}{a}{}{;}

\newcommand{\lm}{$L_{\rm{bol}}/M_{\rm{clump}}$\xspace}
\newcommand{\lsun}{L$_\odot$\xspace}

\newcommand{\msun}{M$_\odot$\xspace}

\newcommand{\lbol}{$L_{\rm{bol}}$}
\newcommand{\mclump}{$M_{\rm{clump}}$\xspace}

\newcommand{\mum}{$\mu$m\xspace}

\newcommand{\kms}{km\,s$^{-1}$\xspace}

\newcommand{\hii}{H\textsc{ii}\xspace}
\newcommand{\co}[2]{CO\,(#1--#2)\xspace}

\newcommand{\rgc}[1]{$R_{\rm gc} = $#1\,kpc}

\newcommand{\pvalue}[1]{$p$-value\,=\,#1}

\newcommand{\higal}{Hi-GAL\xspace}
\newcommand{\oghres}{OGHReS\xspace}
\newcommand{\complexname}{G195.65$-$0.10\xspace}
\newcommand{\complexnamelambda}{$\lambda$-complex\xspace}

\title[The $\lambda$-complex]{OGHReS: Discovery of a giant hub-filament system at the edge of the Galactic molecular disc\thanks{The full version of Table\,\ref{tbl:derived_clump_para} is only available in electronic form at the CDS via anonymous ftp to cdsarc.u-strasbg.fr (130.79.125.5) or via http://cdsweb.u-strasbg.fr/cgi-bin/qcat?J/MNRAS/.}}

\author[J.\,S.\,Urquhart et al.]{J.\,S.\,Urquhart\orcidlink{0000-0002-1605-8050},$^{1}$\thanks{E-mail: j.s.urquhart@kent.ac.uk}, 
A.\,Y.\,Yang\orcidlink{0000-0003-4546-2623},$^{2}$\thanks{E-mail: yangay@nao.cas.cn} 
M.\,A.\,Thompson\orcidlink{0000-0001-5392-909X},$^{3,13}$
C.\,K\"onig\orcidlink{0000-0001-8032-9874},$^{4}$ 
D.\,Colombo\orcidlink{0000-0001-6498-2945},$^{5,4,6,7}$ 
\newauthor
W.-J.\, Kim,$^{4}$ 
A.\,Giannetti,$^{8}$ 
A.\,Karska,$^{9,10}$ 
F.\,Wyrowski,$^{4}$ 
M.\,Figueira,$^{11}$ 
T.\,J.\,T.\,Moore,$^{12}$ 
\newauthor
L.\,K.\,Morgan\orcidlink{0000-0001-8708-5593},$^{14}$ 
Enrique V\'azquez-Semadeni\orcidlink{0000-0002-1424-3543},$^{15}$ 
\\
$^{1}$ Centre for Astrophysics and Planetary Science, University of Kent, Canterbury, CT2\,7NH, UK \\
$^{2}$ State Key Laboratory of Radio Astronomy and Technology, National Astronomical Observatories, Chinese Academy of Sciences,\\ 
\hspace{5pt} A20 Datun Road, Chaoyang District, Beijing, 100101, China\\
$^{3}$ School of Physics and Astronomy, University of Leeds, Leeds LS2 9JT, UK\\
$^{4}$ Max-Planck-Institut f\"ur Radioastronomie (MPIfR), Auf dem H\"ugel 69, 53121 Bonn, Germany\\
$^{5}$ Argelander-Institut f\"ur Astronomie, Universit\"at Bonn, Auf dem H\"ugel 71, 53121 Bonn, Germany \\
$^{6}$ INAF - Osservatorio Astronomico di Cagliari, Via della Scienza 5, 09047 Selargius (CA), Italy\\
$^{7}$ Cluster of Excellence "Our Dynamic Universe" (DYNAVERSE), Institute of Physics I, University of Cologne, Z\"ulpicher Str. 77, 50937 Cologne, Germany\\
$^{8}$ INAF - Istituto di Radioastronomia, Via P. Gobetti 101, I-40129 Bologna, Italy\\
$^{9}$ Institute of Advanced Studies, Nicolaus Copernicus University in Torun, Wileńska 4, 87-100 Toruń, Poland\\
$^{10}$ I Physikalisches Institut, Universität zu K\"oln, Z\"ulpicher Str. 77, D-50937 K\"oln, Germany\\
$^{11}$ National Centre for Nuclear Research, Pasteura 7, 02-093, Warszawa, Poland\\
$^{12}$ Astrophysics Research Institute, Liverpool John Moores University, Liverpool Science Park, 146 Brownlow Hill, Liverpool, L3\,5RF, UK\\
$^{13}$ INAF - Osservatorio Astrofisico di Catania, Via Santa Sof\'ia 78, I-95123 Catania, Italy\\
$^{14}$ NRAO Green Bank Observatory, 155 Observatory Rd, PO Box 2, Green Bank, WV 24944, USA\\
$^{15}$ Instituto de Radioastronom\'ia y Astrof\'isica,
      Universidad Nacional Aut\'onoma de M\'exico, Apdo. postal 3-72, Morelia Mich. 58089, M\'exico 
}

\date{Accepted XXX. Received YYY; in original form ZZZ}

\pubyear{2026}

\begin{document}
\label{firstpage}
\pagerange{\pageref{firstpage}--\pageref{lastpage}}
\maketitle

\begin{abstract}
We present the discovery and characterisation of \complexname (also known as the $\lambda$-complex), a giant hub-filament system identified by the Outer Galaxy High Resolution Survey (OGHReS). The complex contains 61 spatially and kinematically coherent \higal clumps with a mean velocity of $V_{\rm LSR}=32.85$\,\kms and a velocity dispersion between clumps of 1.97\,\kms. A $Gaia$ parallax measurement of the central emission-line star MWC\,137 associated with an embedded cluster gives a heliocentric distance of 5.2$^{+0.8}_{-0.6}$\,kpc, placing the complex at a Galactocentric radius of 13.2\,kpc. The molecular structure spans approximately $35\times63$\,pc and contains $\sim 1.6\times10^{5}$\,\msun of molecular gas, including $\sim 5 \times 10^{4}$\,\msun in dense clumps. The molecular emission is organised into three prominent velocity-coherent filaments that converge on a dense central hub. Of the associated clumps, 16 are protostellar and 45 are gravitationally bound but currently starless. The protostellar clumps are significantly more concentrated towards the hub than the starless population, with median projected offsets of 6.65 and 15.83\,pc, respectively. Moreover, the luminosity-to-mass ratio of the protostellar clumps decreases significantly with increasing distance from the hub, indicating that the most evolutionarily advanced sources are preferentially located in the central region. These results reveal a systematic concentration of dense gas and increasingly evolved star formation towards the hub, with organised large-scale gas motions throughout the filamentary network. The $\lambda$-complex therefore provides an unprecedented laboratory to understand hierarchical and high-mass star formation in the outer Galactic disc.
\end{abstract}

\begin{keywords}
stars: formation – ISM: clouds – ISM: kinematics and dynamics – Galaxy: structure
\end{keywords}



\section{Introduction}
\label{sect:introduction}

Filamentary molecular clouds are now recognised as a fundamental component of the Galactic interstellar medium and play a central role in the star formation process. They are observed over a wide range of spatial scales, from giant molecular filaments (GMFs; \citealt{li2013,ragan2014,ragan2015,Abreu-Vicente2016,colombo2021, Mookerjea2026}) extending for hundreds of parsecs and tracing spiral structure, to the dense filamentary networks embedded within giant molecular clouds (\citealt{Andre2014, inutsuka2015, li2016, mattern2018,Schisano2020}). Some of the largest filamentary structures are associated with major segments of spiral arms and have become known as the so-called ``bones'' of the Milky Way (\citealt{goodman2014,zucker2018}), emphasising the importance of filaments in defining Galactic structure (see \citealt{hacar2023} for a recent review).

Dense filaments are also intimately linked with star formation. Young stellar objects are preferentially found within filamentary structures (e.g. \citealt{lada2012}), while high-mass star formation is frequently concentrated towards dense hubs formed where multiple filaments intersect (\citealt{myers2009, motte2018}). These hub-filament systems (HFSs) exhibit enhanced concentrations of dense gas, embedded clusters, and massive protostars, and observations increasingly suggest that material is funnelled along the filaments towards the central hub, potentially regulating the formation of stellar clusters and high-mass stars (e.g. \citealt{schneider2012, mallick2013, peretto2013,peretto2014, motte2018, williams2018, zhou2022}).

Large catalogues of HFSs have recently become available through systematic surveys of the Galactic plane via $Herschel$ data. \citet{kumar2020} identified many thousands of candidate hub-filament systems in the \higal survey (\citealt{elia2017}), demonstrating that approximately 10\,per cent of dense Galactic clumps reside within these structures. However, despite their abundance, almost all currently known HFSs are located within the inner Galaxy. Indeed, approximately 98.5\,per cent of the Kumar et al. sample lies at Galactocentric radii smaller than the Solar circle (\rgc{8.15}; \citealt{reid2019, GRAVITY2019}). Only 52 candidates are found in the outer Galaxy and these are all located on the far side of the Galactic disc at heliocentric distances exceeding 20\,kpc, making detailed studies of their internal structure extremely difficult. Moreover, these distances are kinematic in nature and therefore likely to be quite uncertain, further complicating the analysis and interpretation of their properties.

The scarcity of well-resolved outer-Galaxy hub-filament systems represents an important limitation for understanding clustered star formation under environmental conditions that differ substantially from those in the inner Milky Way. One of the main reasons for this scarcity is that there has yet to be a systematic search for these systems in the outer Galaxy; \citealt{kumar2020} focused their efforts towards the inner Galaxy \higal catalogue \citep{elia2017} that was available at the time. The outer Galaxy has lower gas surface densities (\citealt{heyer1998_outer}), lower metallicities (\citealt{lepine2011,Esteban2017}), weaker spiral perturbations and reduced stellar densities, all of which may influence the formation, evolution and star-forming properties of filamentary molecular clouds. Consequently, it remains unclear whether the hub-filament paradigm established from studies of nearby and inner-Galaxy regions also applies in these more extreme environments.

In this paper we present the discovery and characterisation of G195.65$-$0.10, a giant hub-filament system identified by the Outer Galaxy High Resolution Survey (OGHReS). Located at a Galactocentric radius of 13.2\,kpc (heliocentric distance of 5.15\,kpc), the complex is consistent with being associated with the Outer-Norma arm and is one of the most distant well-resolved hub-filament systems currently known. Spanning approximately $35\times63$\,pc, it is considerably larger than the majority of previously studied HFSs, which typically extend over only $10$--$20$\,pc (e.g. \citealt{kumar2020}; determined from the total length of filaments outside of the central hub), while its relatively nearby heliocentric distance provides a physical resolution several times better than that achieved for currently known outer-Galaxy examples. These characteristics make G195.65$-$0.10 an exceptional laboratory in which to investigate the relationship between filamentary gas structure and kinematics, clustered and high-mass star formation, and the large-scale environment of the outer Milky Way.

The structure of this paper is as follows. In Sect.\,\ref{sect:HFS_identification}, we briefly describe the OGHReS survey and introduce G195.65$-$0.10. Section\,\ref{sect:G195_properties} presents the physical properties of the complex, its molecular gas distribution and star formation characteristics. In Sect.\,\ref{sect:discussion}, we discuss the implications of these results and compare G195.65$-$0.10 with other hub--filament systems reported in the literature before summarising our conclusions in Sect.\,\ref{sect:conclusions}.

\section{Discovery of the $\lambda$-Complex}
\label{sect:HFS_identification}

 \begin{figure*}
\centering
        
        \includegraphics[width=0.80\textwidth]{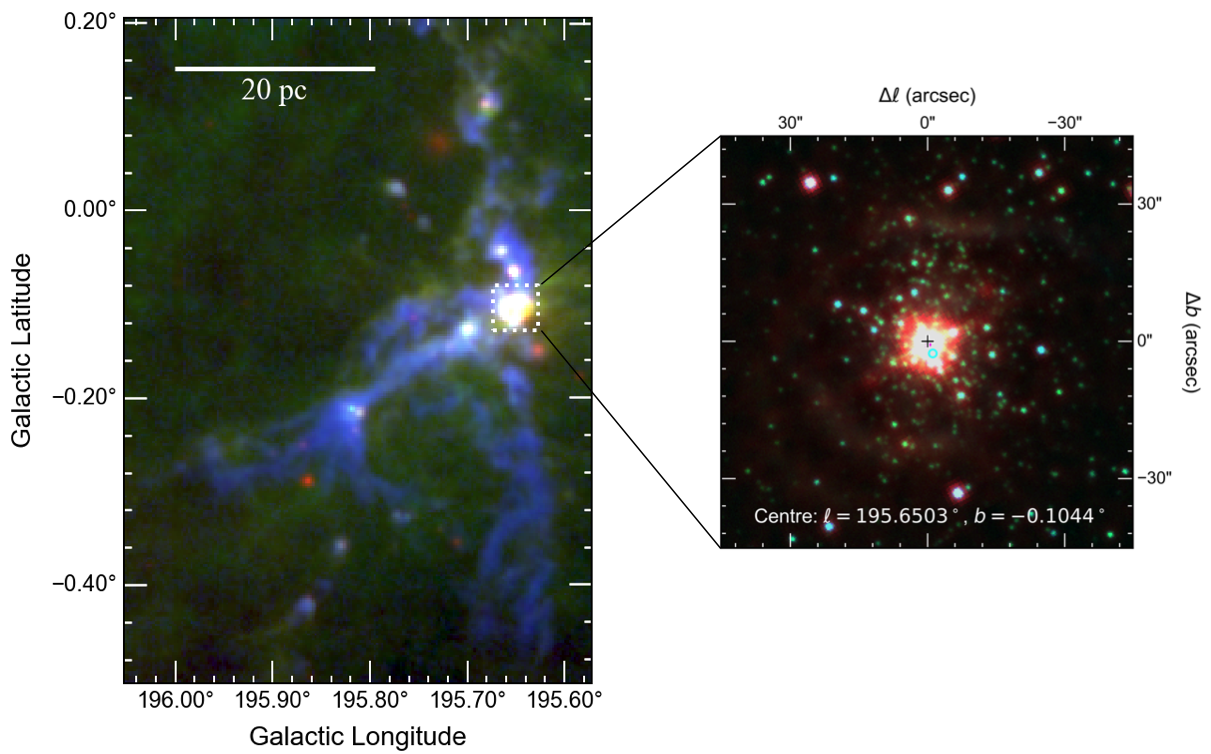}
    
   	\caption{
    Three-colour composite images of the hub-filament system \complexname. The image in the left panel has been constructed using the WISE 22\,\mum (red), the \higal\ 160\,\mum (green) and integrated CO emission (blue); these wavelengths have been selected to contrast the distribution of emission from hot gas, cooler dust emission and cold molecular gas, respectively. The image shown in the right panel is a zoom of the central hub of this complex and is constructed from the UKIDSS $J$  and $H$ band filters (\citealt{lawrence2007}; blue and green, respectively) and the 2MASS $K_s$ filter (\citealt{2mass}; red).
    }
	\label{fig:complex_map}
	
\end{figure*}

\subsection{OGHReS and the identification of molecular complexes}
\label{sect:oghres_survey}

The Outer Galaxy High Resolution Survey (OGHReS; {\color{blue}K\"onig et al. in prep.}) is a systematic high-resolution survey of \co{2}{1} emission across approximately 100\,deg$^2$ of the third Galactic quadrant ($180\degree < \ell < 280\degree$ and 1\degr\ in $b$), obtained using the Atacama Pathfinder EXperiment (APEX; \citealt{gusten2006}) 12~m telescope. The survey provides molecular-line data with an angular resolution of approximately 30\,arcsec and a typical sensitivity of $T_{\rm mb}\simeq0.4$\,K per 0.25\,\kms channel, allowing the molecular cloud population of the outer Galaxy to be investigated at a level of detail previously only possible in the inner Galaxy.

In two recent papers, \citet[][hereafter Papers\,I and II]{urquhart2024, urquhart2025}, the OGHReS data were used to verify and, where necessary, update the velocities of $\sim$6\,200 \higal sources (\citealt{elia2021}). These velocity measurements were subsequently combined with the projected positions of the clumps in a minimum spanning tree (MST) analysis (see \citealt{NESETRIL20013} for a review and \citealt{wang2016} for an application to Galactic star-forming regions).  Maximum linking lengths of 6\,arcmin and 2\,\kms were adopted in angular position and velocity, respectively, allowing groups of dense clumps that are both spatially and kinematically coherent to be robustly identified. These groups correspond to molecular complexes and filamentary structures that are likely to represent physically connected star-forming regions.

This analysis has identified hundreds of molecular complexes and filamentary structures and the full catalogue will be presented in a forthcoming paper. During inspection of the largest and most significant structures, one complex immediately stood out owing to its exceptional size, organised hub-filament morphology and rich population of dense clumps. This object, centred at $\ell=195.65^\circ$ and $b=-0.10^\circ$, is the focus of the present paper.

\subsection{The \complexname Complex}
\label{sect:g195_complex}

The MST analysis identified a coherent group of 61 \higal clumps centred at $\ell=195.72^\circ$, $b=-0.13^\circ$, with a mean velocity of $V_{\rm LSR}=32.85$\,\kms. The associated clumps have angular dispersions of $\Delta\ell=4.7$\arcmin\ and $\Delta b=9.4$\arcmin, together with an overall velocity dispersion of $\Delta \upsilon=1.97$\,km\,s$^{-1}$, indicating that they form a compact, kinematically coherent structure. Inspection of the OGHReS $^{12}$CO emission integrated over the same velocity range as the clump reveals that the clumps are embedded within a giant hub-filament system with an extended reservoir of molecular gas that is both spatially and kinematically coherent, confirming that they are part of a single star-forming complex.

The left panel of Figure\,\ref{fig:complex_map} presents a three-colour image combining the integrated $^{12}$CO (2--1), \textit{Herschel} 160\,$\mu$m and $WISE$ 22\,$\mu$m emission. These wavelengths have been chosen to emphasise the distribution of cold molecular gas, cooler dust emission and warm dust associated with embedded star formation. This image reveals the complex to be a highly structured hub-filament system exhibiting significant ongoing star-formation activity. The molecular gas resides in three prominent filaments extending towards the north, south and south-east that converge on a dense central hub. Compact infrared sources are distributed along the filaments, while the hub itself is associated with bright, extended mid-infrared emission indicative of ongoing high-mass star formation. 

The central hub is shown in greater detail in the right-hand panel of Fig.\,\ref{fig:complex_map}, which presents a three-colour near-infrared image constructed from the UKIDSS $J$ and $H$ bands (\citealt{lawrence2007}) together with the 2MASS $K_s$ band (\citealt{2mass}). This image reveals a compact cluster of bright stars surrounded by diffuse infrared emission offset by approximately 20\,arcsec from the geometric centre of the hub. The region has previously been identified with the IRAS source IRAS\,06158+1517 ($\ell=195.6502^\circ$, $b=-0.1046^\circ$), which is associated with the peculiar emission-line star MWC\,137
(\citealt{Kraus2021}). MWC\,137 has been classified as a Herbig B[e] supergiant and is surrounded by the optical nebula Sh\,2-266 and an embedded infrared cluster (\citealt{bica2003a}).

The central hub has also been identified as the H\,{\sc ii} region G195.648$-$00.106 (\citealt{Anderson2015}) from analysis of WISE data (\citealt{Wright2010}). Radio recombination line observations measure a velocity of approximately 28\,km\,s$^{-1}$ (\citealt{Anderson2015}), which is consistent with both the molecular gas (30.13\,\kms; \citealt{Wouterloot1989}) and the velocity of the associated Hi-GAL clumps. A search of the NVSS catalogue (\citealt{condon1998}) reveals a coincident 20-cm radio continuum source, NVSS\,J061845+151652, with an integrated flux density of $93.5\pm3.5$\,mJy. At a distance of $\sim$5\,kpc (see Sect.\,\ref{sect:distance_size} for details),  this corresponds to a Lyman continuum photon flux of $\log(N_{\rm Ly})=47.36$ (\citealt{carpenter1990}), consistent with ionisation by an  early B-type star corresponding to a $\sim15\,{\rm M_\odot}$ zero-age main-sequence (ZAMS) star
(\citealt{davies2011}).

Although the central hub has been studied previously, its association with the extensive molecular filamentary network revealed by OGHReS has not been recognised. We designate the complete structure \complexname after the Galactic coordinates of its central hub; however, given its morphological similarity to the Greek letter $\lambda$, we also refer to this object as the $\lambda$-complex.

\section{Complex Properties}
\label{sect:G195_properties}

\begin{figure}
	\centering
        \includegraphics[width=0.49\textwidth, trim = 0cm 0.5cm 0cm 0cm, clip]{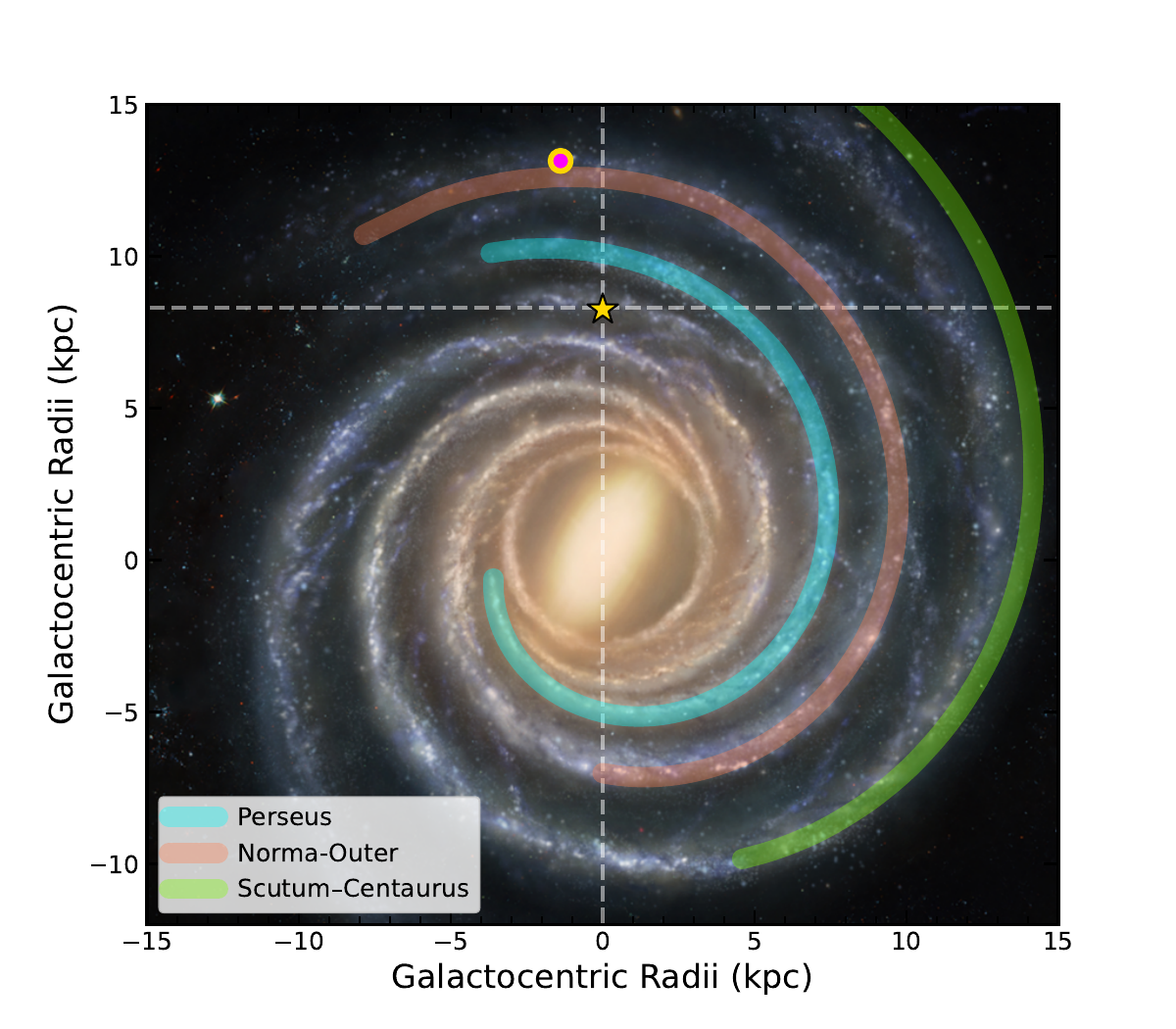}
        
    	\caption{Top-down view of the Milky Way showing the  position of the complex determined from the $Gaia$ parallax data (filled red circle). The background image is a schematic representation of the Milky Way based on approximately 200 maser parallax measurements (credit: Xing-Wu Zheng \& Mark Reid, BeSSeL/NJU/CfA). We highlight the most relevant segments of Perseus, Norma (\citealt{reid2019}) and Scutum-Centaurus (\citealt{sun2015,sun2024}) spiral arms in cyan, red and green respectively. The position of the Sun is shown by the yellow star.
        }
		\label{fig:xy_diagrams}
	
\end{figure}

\subsection{Distance, mass, size and Galactic location}
\label{sect:distance_size}

Using the molecular velocity of 30.13\,\kms\ measured towards the central IRAS source and the Galactic rotation curve of \citet{brand1993}, \citet{Wouterloot1989} derived Galactocentric and heliocentric distances of \rgc{17.76} and $D=9.42$\,kpc, respectively. Repeating this calculation using the mean velocity of the associated \higal\ clumps ($V_{\rm LSR}=32.85$\,\kms) together with the updated Galactic rotation parameters of \citet{reid2019} ($R_0=8.15$\,kpc and $\Theta_0=240$\,\kms) gives a similar value of \rgc{16.46} and $D=8.47$\,kpc. However, kinematic distances are known to be highly uncertain towards the Galactic anti-centre owing to the shallow velocity gradient in this direction and so these values should be treated with caution.

 \begin{figure*}

\centering
        
    \includegraphics[width=0.45\textwidth]{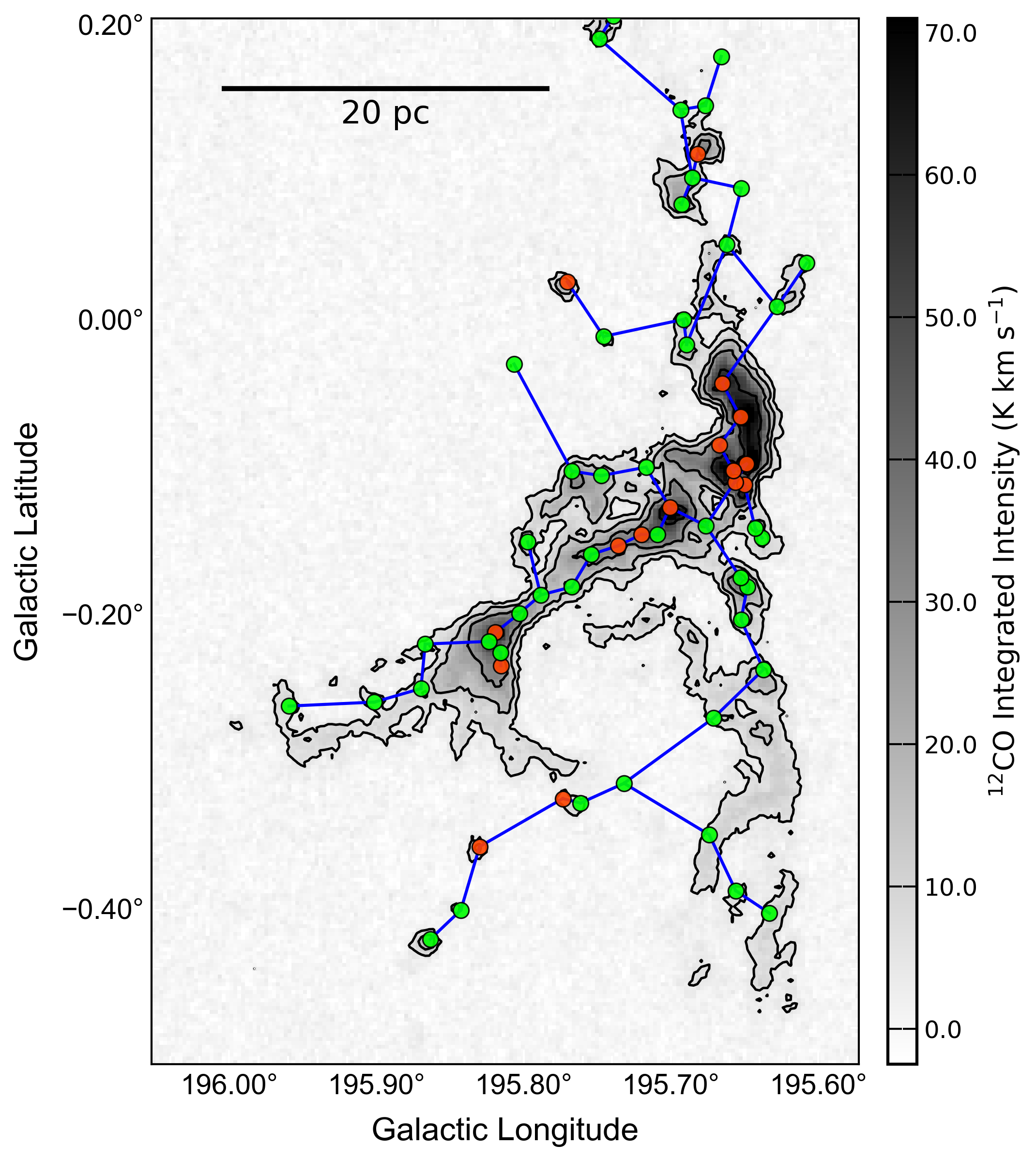}
    \includegraphics[width=0.45\textwidth]{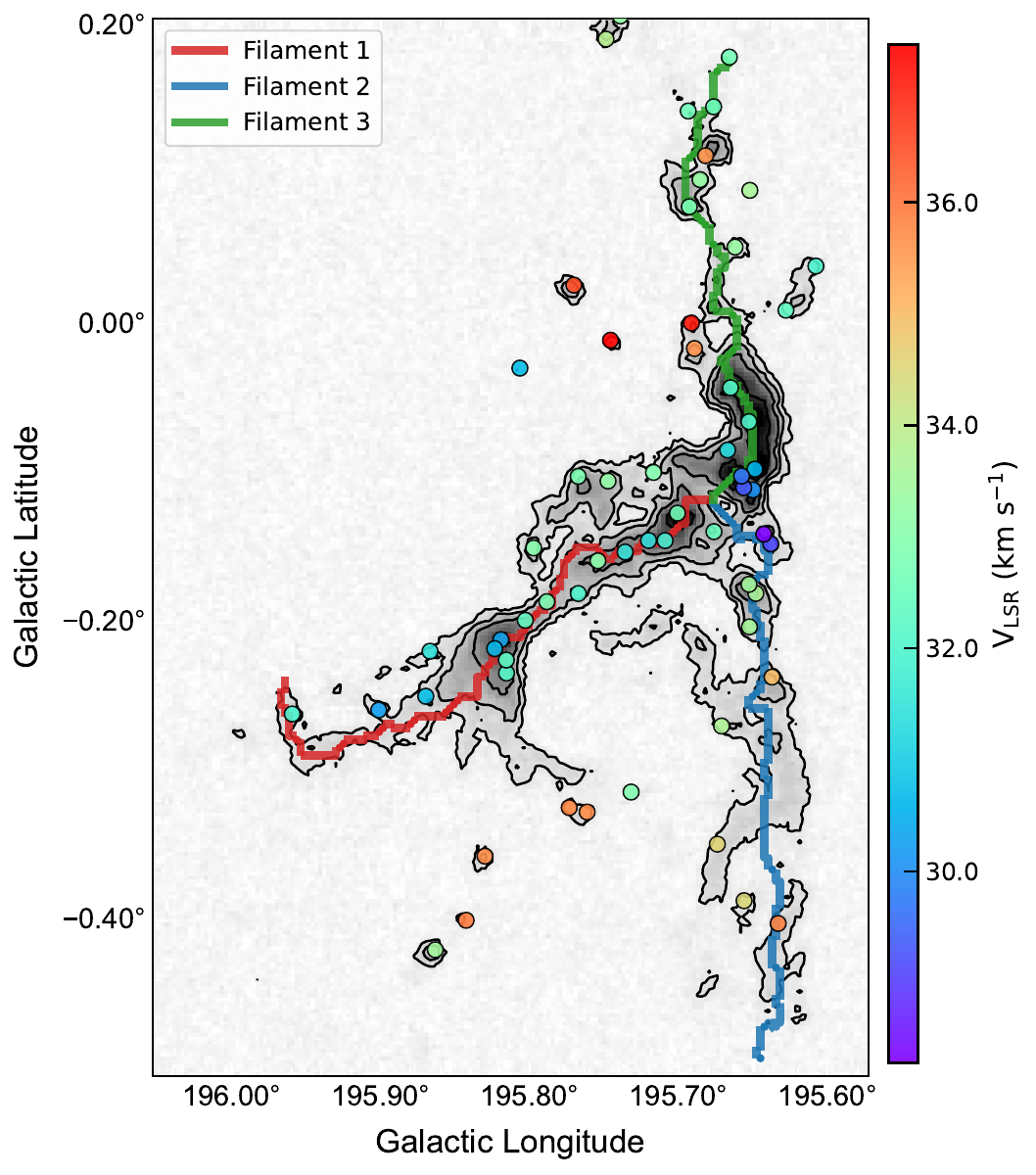}\\
    \includegraphics[width=0.45\textwidth]{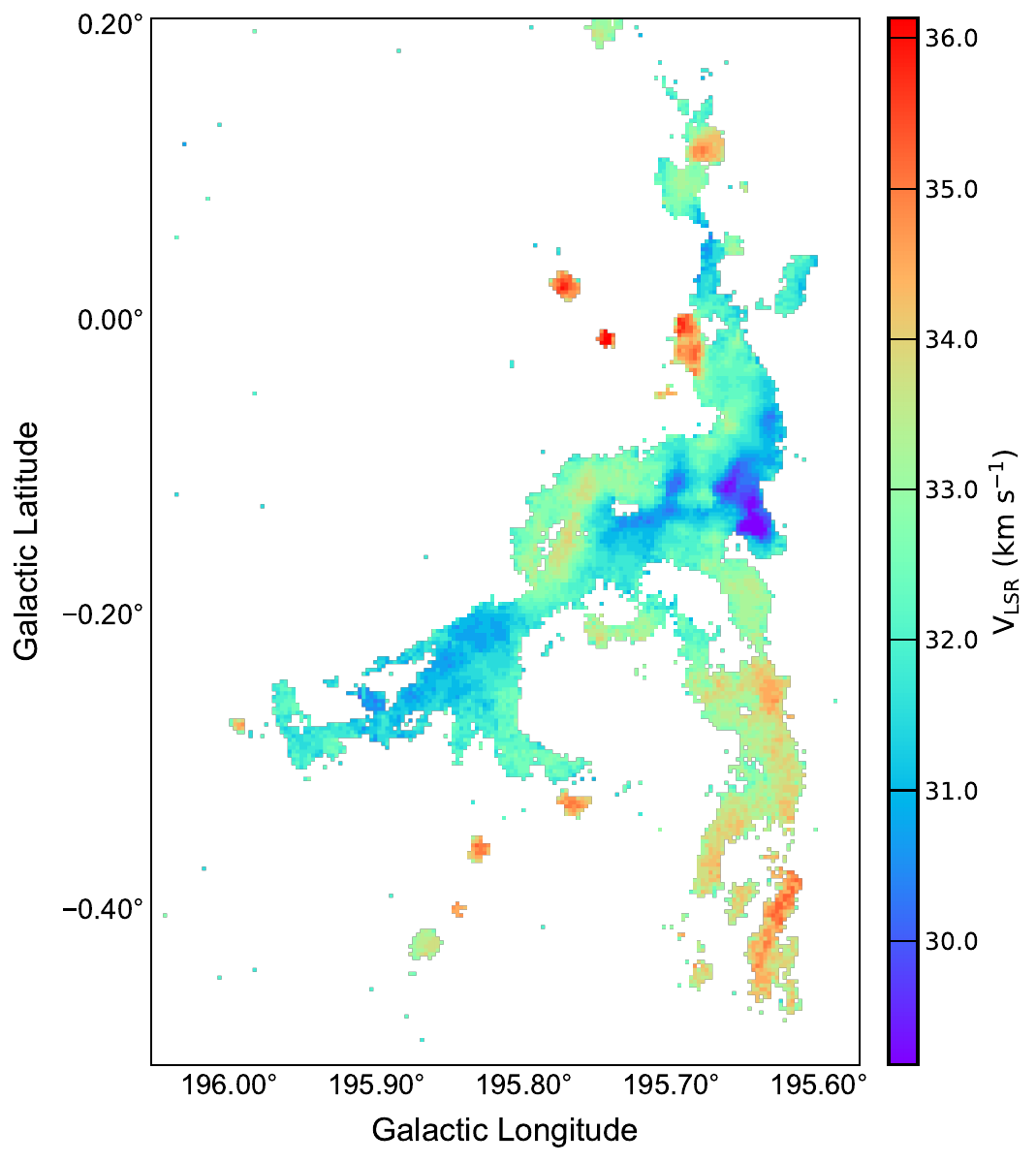}
    \includegraphics[width=0.45\textwidth]{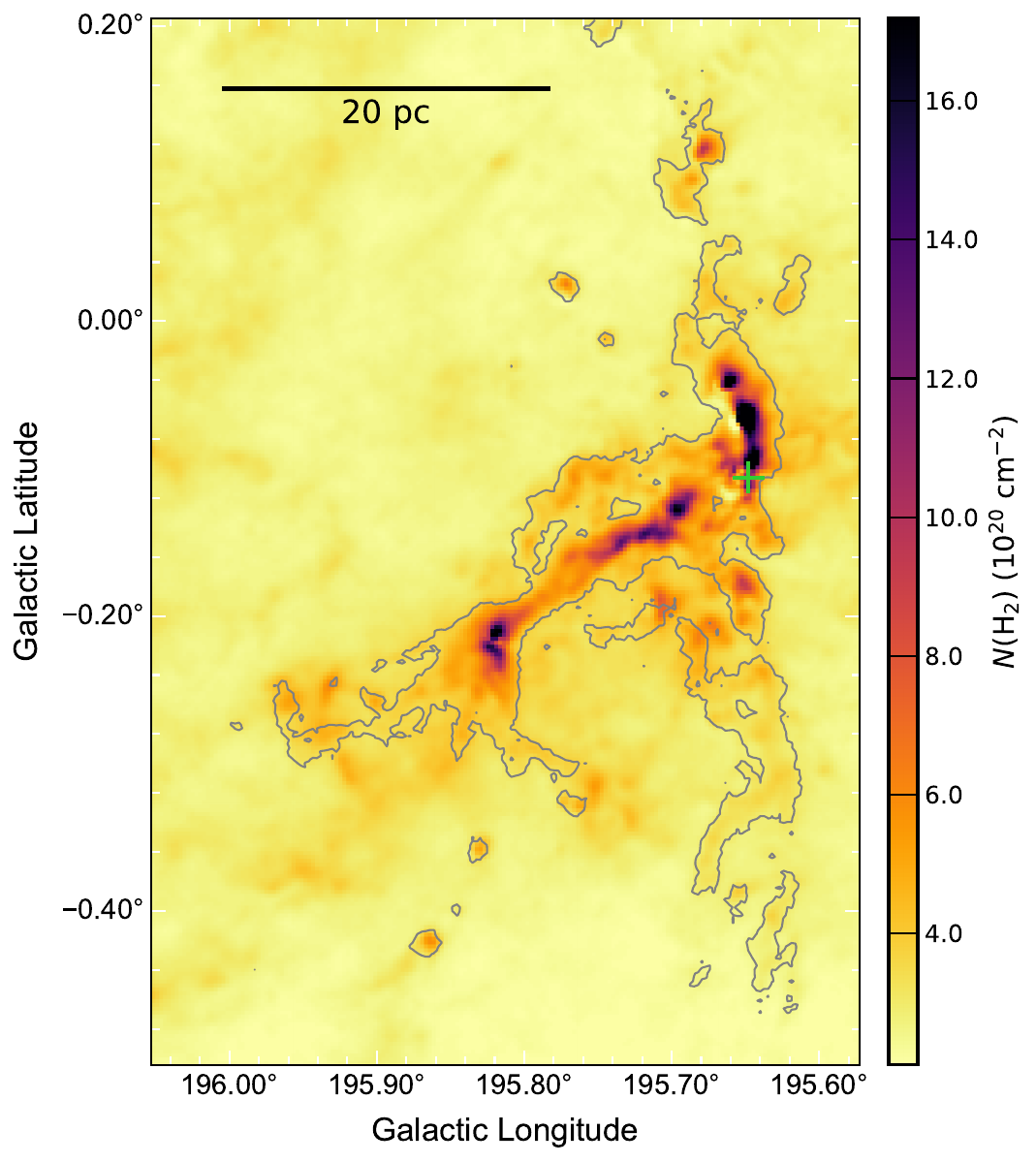}

   	\caption{Maps of the \complexnamelambda showing the distribution of different properties or features discussed in the text. The upper panels show the $^{12}$CO moment 0 map of the complex (greyscale) with contours starting at 2$\sigma$ and increasing in steps of 5$\sigma$, 10$\sigma$, 20$\sigma$ and 30$\sigma$. In the upper left panel, we show the positions of the \higal clumps that are both kinematically and spatially coincident with the complex; these are coloured according to evolutionary stage (red and green filled circles correspond to protostellar and quiescent bound clumps). The blue connecting lines show the results of the MST analysis. The coloured filled circles shown on the upper right panel also show the positions of associated \higal\ clumps but here the colours are linked to the clump velocities (see colour bar for corresponding values). The red, blue and green lines trace the spines of the filaments identified (see Sect.\,\ref{sect:filaments} for details). In the lower left panel, we show the  $^{12}$CO moment 1 map, which shows the intensity-weighted mean velocity across the complex. In the lower right panel, we show the column density map produced by \citet{marsh2015} from the \higal\ continuum maps; the contour traces the 2$\sigma$ level of the integrated $^{12}$CO emission shown in the upper panels of this figure.  }
	\label{fig:clump_properties}
	
\end{figure*}

Fortunately, the central hub contains the emission-line star MWC\,137, which is associated with the \textit{Gaia} DR3 source 3344973478481675264. The stellar parallax of this source ($0.194\pm0.0256$\,mas) corresponds to a heliocentric distance of $D=5.2^{+0.8}_{-0.6}$\,kpc and a Galactocentric distance of \rgc{13.2} \citep{Kraus2021}.  At this distance, we would expect the complex to have a metallicity approximately 0.4\,$Z_\odot$ assuming the radial metallicity gradient (\citealt{Esteban2017}) and therefore provides a window to study star formation in a very different environment and a link to extragalactic studies.  Given the close spatial coincidence between MWC\,137, the compact H\,{\sc ii} region and the centre of the molecular hub, we adopt the \textit{Gaia}-based distance for the entire complex. This distance is consistent with the \complexnamelambda being associated with the Outer-Norma spiral arm as shown in Fig.\,\ref{fig:xy_diagrams}. This also places this complex towards the edge of the Galactic molecular disc (\rgc{13.5}; \citealt{heyer1998_outer}), making it the most distant high-mass star-forming complex yet discovered with respect to the Galactic centre. Examination of mid-infrared and NVSS radio maps of the extended environment does not reveal any larger scale external features that might have influenced the formation and subsequent evolution of this complex. 

The total molecular gas mass of the \complexnamelambda is estimated independently from the integrated $^{12}$\co{2}{1} emission and the dust-derived H$_2$ column-density map of \citet{marsh2015}. The spatial extent of the complex was defined using the 2$\sigma$ contour of the $^{12}$CO velocity-integrated intensity map. This mask was applied to both the integrated CO emission and the corresponding \higal H$_2$ column-density map to identify the emission associated with the complex (see upper and lower right panels of Fig.\,\ref{fig:clump_properties}, respectively).

\begin{figure*}
	\centering
        \includegraphics[width=0.85\textwidth, trim = 0cm 0cm 0cm 0cm, clip]{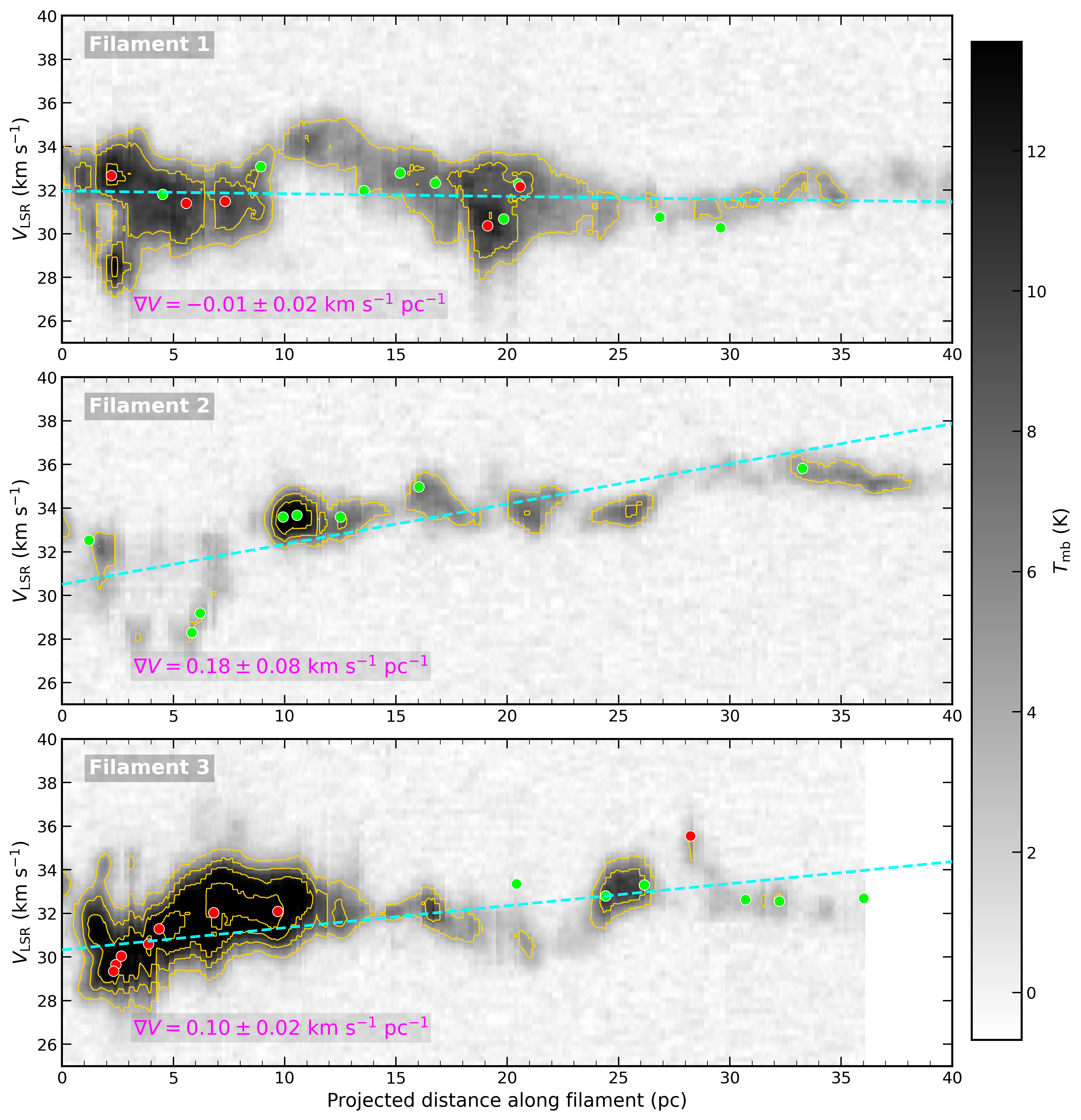}
           
    	\caption{$^{12}$CO position-velocity diagrams for the three filaments. The position–velocity cuts begin where the filaments converge and follow their spines (see upper right panel of Fig.\,\ref{fig:clump_properties} for details). The coloured circles indicate the position of protostellar (red) and starless (green) \higal clumps associated with each filament; which we define as being within a projected distance of 1.5\,pc of the filament. The dashed cyan line shows the result of a linear fit to the molecular emission;  the fitted velocity gradients are given in the lower-left corner of each panel.}
		\label{fig:pv_diagrams}
	
\end{figure*}

For the CO data, H$_2$ column densities were derived from the integrated $^{12}$\co{2}{1} intensities, $I_{\rm CO\,(2-1)}$, measured from the \oghres moment-0 map, using:

    \begin{equation}
N({\rm H_2}) =
X_{\rm CO(1-0)}
I_{\rm CO(2-1)}
R_{21}^{-1}
\left(\frac{Z}{Z_\odot}\right)^{-1.5},
\end{equation}

\noindent where $X_{\rm CO(1-0)}=2.3\times10^{20}$\,cm$^{-2}$ (K\,km\,s$^{-1}$)$^{-1}$ (\citealt{brand1995}), $R_{21}$ is the \co{1}{0} to \co{2}{1} line ratio taken as $0.7$ (\citealt{Sandstrom2013, Nishimura2015}) and $\left(\frac{Z}{Z_\odot}\right)^{-1.5}$ is the correction for the CO-dark gas fraction (\citealt{Schinnerer2024}). The resulting column-density map was integrated over the complex mask assuming a mean molecular weight of $\mu_{\rm H_2}=2.8$ to give the total mass of the complex of $1.65\times 10^5$\,\msun.

The same spatial mask was applied independently to the dust-derived H$_2$ column-density map. This gives a total dust mass of the complex as $M_{\rm dust}=4.0\times10^4$\,\msun assuming a standard local gas-to-dust ratio of 100 \citep{elia2021}. However, at the distance of this complex from the Galactic centre (\rgc{13.2}) the gas-to-dust ratio is expected to be substantially larger than the local value, reaching approximately 400 \citep{giannetti2017_g2d} at the distance of this complex. Applying the appropriate Galactocentric-dependent correction increases the estimated total dust derived gas mass to approximately $1.6\times10^5\,{\rm M}_{\odot}$, which is in excellent agreement with the total mass derived from the CO analysis presented above.

The complex extends approximately 0.4\degr\ in Galactic longitude and 0.7\degr\ in latitude\footnote{To put this into context, the size of this complex is comparable to the angular size of the full Moon.}, corresponding to projected physical dimensions of approximately $35\times63$\,pc at the adopted distance of 5.2\,kpc. For comparison, most previously identified hub-filament systems have characteristic sizes of approximately 10–20\,pc \citep{kumar2020}. The \complexnamelambda is therefore both substantially larger than a typical HFS and located at a large Galactocentric radius. Its heliocentric distance also provides a physical resolution of approximately 0.75\,pc for the 30-arcsec OGHReS beam, considerably better than that available for the most distant outer-Galaxy HFS candidates.

\subsection{Filamentary structure and kinematics}
\label{sect:filaments}

\begin{table}
\setlength{\tabcolsep}{5pt}
\begin{center}\caption{Derived filament properties.}
\label{tbl:derived_filament_para}
\begin{minipage}{\linewidth}

\begin{tabular}{c...cc}
\hline \hline
  \multicolumn{1}{c}{Filament}& 
  \multicolumn{1}{c}{Length}&
  \multicolumn{1}{c}{Width}&
  \multicolumn{1}{c}{Aspect}&
  \multicolumn{1}{c}{Velocity gradient} &
  \multicolumn{1}{c}{$\Delta V$} 
   \\

    \multicolumn{1}{c}{designation}&  
    \multicolumn{1}{c}{(pc)}&
    \multicolumn{1}{c}{(pc)}&
    \multicolumn{1}{c}{ratio} & 
     \multicolumn{1}{c}{(\kms\,pc$^{-1}$)}&
     \multicolumn{1}{c}{(\kms)}\\

\hline
1 & 41.1&3.1&13.2&$-0.01\pm0.02$ & $\cdots$ \\
2 & 41.8 &2.1&19.7&$0.18\pm 0.08$ & 7.5\\
3&33.6&2.2&14.9&$0.10\pm0.02$ & 3.4 \\

\hline
\end{tabular}\\

\end{minipage}

\end{center}
\end{table}
\setlength{\tabcolsep}{6pt}

The morphology of the molecular gas complex is dominated by three extended filamentary structures that converge on the central hub. To quantify their morphology and kinematics, we developed a graph-based analysis combining dendrogram segmentation, skeletonisation and network analysis. The molecular complex was first isolated from the surrounding diffuse emission and reduced to a one-pixel-wide skeleton. The resulting graph representation was then used to identify the principal filamentary spines extending from the central junction and to determine their lengths, widths and aspect ratios.

The three principal structures are shown in the upper right panel of Fig.\,\ref{fig:clump_properties} and are labelled Filaments\,1, 2 and 3 (coloured red, blue and green, respectively), and their derived properties are listed in Table\,\ref{tbl:derived_filament_para}. They have projected lengths of 41.1, 41.8 and 33.6\,pc and characteristic widths of 3.1, 2.1 and 2.2\,pc, respectively. Their corresponding aspect ratios range from 13.2 to 19.7, demonstrating that all three structures are highly elongated.

Both the associated \higal clump velocities and the 
$^{12}$CO moment-1 map reveal a coherent large-scale velocity pattern across the complex (as evidenced by the individual clumps velocities and moment-1 map of the CO gas shown in the upper right and lower left panels of Fig.\,\ref{fig:clump_properties}), with the lowest velocities generally associated with the central hub and increasing velocities towards larger projected distances along the filaments. Superposed on this large-scale pattern are smaller-scale variations associated with individual dense condensations.

To investigate these trends quantitatively, position-velocity cuts were extracted along the three filament spines, beginning at their intersection in the central hub (these are shown in Fig.\,\ref{fig:pv_diagrams}). Linear least-squares fits were used to characterise the variation in velocity with projected distance. The fitted gradients are $-0.01\pm0.02$, $0.18\pm0.08$, and $0.10\pm0.02$\,\kms pc$^{-1}$ for Filaments 1, 2 and 3, respectively. These values lie comfortably within the range typically reported for giant molecular filaments (e.g. \citealt{ragan2014}). The position-velocity diagrams also show the positions and velocities of the associated \higal clumps lying within a projected distance of 1.5\,pc of each filament spine.

The molecular emission and dense-clump velocities are closely correlated along the filamentary structures. The position-velocity diagrams additionally demonstrate that the protostellar clumps preferentially coincide with the highest-intensity molecular emission, consistent with their association with localised high-column-density regions. Intriguingly, we note the presence of oscillations in the velocity distribution, which have been reported in other regions (e.g., \citealt{Henshaw2020, Neupane2024,  he2026}), however, detailed analysis of these features is beyond the scope of the current work.

\subsection{Dense clumps and star formation}

\setlength{\tabcolsep}{3pt}

\begin{table*}


\begin{center}\caption{Derived clump parameters assuming a heliocentric distance of 5.2\,kpc and gas-to-dust ratio of 388. }
\label{tbl:derived_clump_para}
\begin{minipage}{\linewidth}
\small
\begin{tabular}{l..c.........}
\hline \hline
  \multicolumn{1}{c}{\higal}& 
  \multicolumn{1}{c}{$\ell$}&
  \multicolumn{1}{c}{$b$}&
  \multicolumn{1}{c}{Evolution}&
  \multicolumn{1}{c}{T$_{\rm{dust}}$} & 
  \multicolumn{1}{c}{$V_{\rm{LSR}}$}   &
  \multicolumn{1}{c}{$\Sigma$ } &
  \multicolumn{1}{c}{log$_{10}$[$L_{\rm{bol}}$]}	&
  \multicolumn{1}{c}{log$_{10}$[$M_{\rm clump}$]} & 
    \multicolumn{1}{c}{\lm-ratio} \\

    \multicolumn{1}{c}{name}&  
    \multicolumn{1}{c}{(\degr)}&
    \multicolumn{1}{c}{(\degr)}&
    \multicolumn{1}{c}{type} & 
    \multicolumn{1}{c}{(K)}  &
     \multicolumn{1}{c}{(km\,s$^{-1}$)}&	
     \multicolumn{1}{c}{(g\,cm$^{-2}$)} &
    \multicolumn{1}{c}{(\lsun)} &
    \multicolumn{1}{c}{(\msun)} &
      \multicolumn{1}{c}{(\lsun/\msun)} \\
    
\hline

HIGALBM195.6079+0.0392	&	195.6079	&	0.0392	&	Starless	&	11.0	&	31.94	&	0.088	&	1.858	&	0.424	&	0.037	\\
HIGALBM195.6280+0.0096	&	195.6280	&	0.0096	&	Starless	&	11.3	&	32.35	&	0.132	&	2.475	&	1.202	&	0.053	\\
HIGALBM195.6332$-$0.4021	&	195.6332	&	-0.4021	&	Starless	&	10.0	&	35.82	&	0.289	&	2.189	&	0.424	&	0.017	\\
HIGALBM195.6373$-$0.2367	&	195.6373	&	-0.2367	&	Starless	&	9.1	&	34.97	&	0.378	&	2.703	&	0.901	&	0.016	\\
HIGALBM195.6382$-$0.1473	&	195.6382	&	-0.1473	&	Starless	&	8.3	&	29.18	&	0.596	&	3.481	&	1.424	&	0.009	\\
HIGALBM195.6429$-$0.1407	&	195.6429	&	-0.1407	&	Starless	&	14.3	&	28.29	&	0.028	&	1.490	&	0.725	&	0.172	\\
HIGALBM195.6482$-$0.1806	&	195.6482	&	-0.1806	&	Starless	&	11.0	&	33.67	&	0.358	&	2.625	&	1.269	&	0.044	\\
HIGALBM195.6489$-$0.0972	&	195.6488	&	-0.0972	&	Protostellar	&	15.8	&	30.60	&	0.615	&	3.058	&	3.378	&	2.092	\\
HIGALBM195.6501$-$0.1111	&	195.6501	&	-0.1111	&	Protostellar	&	14.6	&	30.04	&	1.059	&	3.238	&	3.317	&	1.200	\\
HIGALBM195.6523+0.0900	&	195.6523	&	0.0900	&	Starless	&	13.9	&	33.62	&	0.047	&	2.268	&	1.503	&	0.172	\\

\hline\\
\end{tabular}\\
Notes: Only a small portion of the data is provided here. The full table is available in electronic form at the CDS via anonymous ftp to cdsarc.u-strasbg.fr (130.79.125.5) or via http://cdsweb.u-strasbg.fr/cgi-bin/qcat?J/MNRAS/.
\end{minipage}

\end{center}

\end{table*}
\setlength{\tabcolsep}{6pt}

The upper-left panel of Fig.\,\ref{fig:clump_properties} shows the distribution of the associated \higal clumps over the integrated $^{12}$\co{2}{1} emission. Protostellar and gravitationally bound starless clumps are indicated by red and green symbols, respectively, while the branches of the MST used to identify the complex are shown in blue. The MST analysis identified 61 \higal clumps associated spatially and kinematically with the \complexnamelambda. Of these, 16 are classified as protostellar, through their association with a 70-$\mu$m point source, while the remaining 45 are classified as gravitationally bound starless clumps (\citealt{elia2021}; \citealt{urquhart2025}). The resulting star formation fraction, defined as the fraction of \higal sources with a 70\,\mum counterpart ($SFF = N_{\rm proto}/N_{\rm all}$; \citealt{ragan2016}), is therefore 0.26. This value is similar to that found for star-forming complexes located in the inner Galactic disc (0.25; \citealt{ragan2016}), indicating that the star formation activity in this complex is comparable with inner Galaxy complexes. 

Although the association of a 70\,\mum counterpart signifies the presence of an embedded protostellar object, the absence of one does not establish that a clump is starless: a lower-mass or less evolved protostar may be too faint to be detected at this wavelength. Approximately 16\,per\,cent of ATLASGAL clumps classified as starless have been identified as outflow candidates (\citealt{urquhart2022, yang2018}). If this fraction is representative of the number of starless clumps in the \higal catalogue, some 70\,\mum-dark clumps may host protostellar objects, rendering the starless classification less definitive. Molecular outflows are among the earliest signposts of star-formation activity \citep[][see their figure 14]{urquhart2022}; an outflow associated with a 70\,\mum-dark clump is therefore consistent with the embedded protostars being at an earlier evolutionary stage than the  70\,\mum-bright population. We use the \higal classifications as useful observational categories, while recognizing that some clumps classified as starless may host early-stage protostars. Although this introduces some uncertainty into the starless sample, the potential number of clumps hosting 70\,\mum-dark protostars is relatively modest and we do not expect this to significantly affect our results or conclusions.

The physical properties of the clumps (\citealt{urquhart2025}) have been rescaled using the adopted $Gaia$ distance to the complex and the appropriate gas-to-dust ratio at its Galactocentric radius (\citealt{giannetti2017_g2d}). The revised surface densities, masses, bolometric luminosities and luminosity-to-mass ratios are listed in Table\,\ref{tbl:derived_clump_para}, together with the clump positions, evolutionary classifications, dust temperatures and molecular velocities.

\begin{figure}
	\centering
         \includegraphics[width=0.45\textwidth, trim = 0cm 0cm 0cm 0cm, clip]{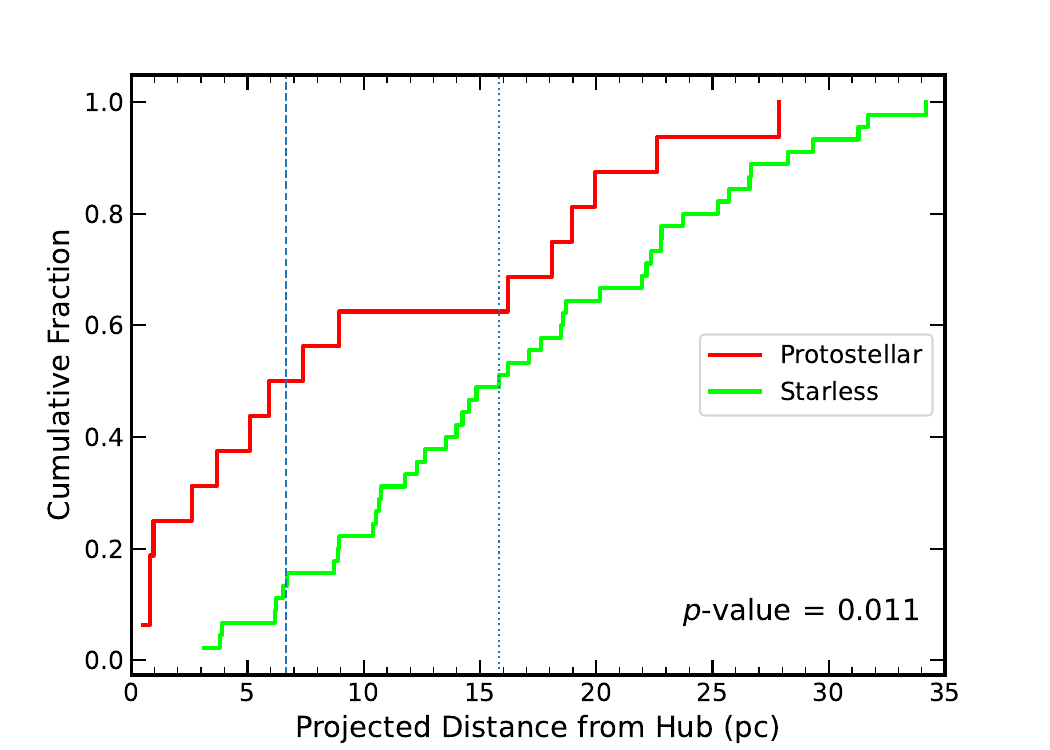}
       
        \caption{Cumulative distribution function showing the projected distance between protostellar and starless clumps with respect to the centre of the hub. The dotted and dashed vertical lines indicate the median projected distance from the hub of the protostellar and starless clumps respectively. A two-sided Mann-Whitney U test gives \pvalue{0.011}, indicating that the protostellar and starless clumps have significantly different projected radial distributions.}
		\label{fig:surface_density}
	
\end{figure}

Summing the individual clump masses gives a total dense-clump mass of approximately $5.0\times10^4\,{\rm M}_{\odot}$. Comparing this with the total molecular mass of the complex gives a dense-gas fraction of approximately 0.3. This value is larger than the $\sim0.1$ reported in several previous studies of inner Galaxy clump samples (e.g. \citealt{eden2015,urquhart2021}), however, these studies used the ATLASGAL compact-source catalogue (\citealt{contreras2013}), which has a different sensitivity to extended dust emission than \higal. A direct quantitative comparison should therefore be treated with caution.

Figure\,\ref{fig:surface_density} shows the cumulative distribution of the projected distance between the protostellar and starless clumps with respect to the central hub. The two populations have noticeably different spatial distributions. Of the 16 protostellar clumps, 10 (63 per cent) are located within 10\,pc of the central hub, compared with 24\,per\,cent of the starless population. The most luminous protostellar clumps are also associated with the hub, with the two most luminous sources having $L_{\rm bol} \approx 2\times 10^3$\,\lsun, while the total bolometric luminosity obtained by summing the luminosities of the embedded \higal clumps is approximately $1.2\times10^4\,{\rm L}_{\odot}$.

The complex contains a substantial population of massive dense clumps, with 18 of the 61 associated Hi-GAL sources having masses exceeding $10^3$\,\msun; nine of which are protostellar. Several of the most massive clumps are concentrated within the high-column-density region along Filament\,1 and have relatively low $L_{\rm bol}/M_{\rm clump}$, indicating that they are at an early evolutionary stage. A clump mass of $10^3$\,\msun is considered to be the threshold for high-mass star formation\footnote{This is determined using the relationship between the most massive star in a cluster compared to the embedded cluster mass (i.e. the $m_{\rm max}- M_{\rm ecl}$; \citealt{Weidner2006,Weidner2010}) and assuming a typical embedded-cluster star formation efficiency of 30\,per\,cent \citep{lada2003,wells2022}.}, and the presence of a significant number of clumps with masses above this threshold may be indicative of ongoing or future high-mass star formation in this complex.

\section{Discussion}
\label{sect:discussion}

 \begin{figure*}

\centering

    \begin{minipage}[b]{0.45\textwidth}
        \centering
        \includegraphics[width=\linewidth]{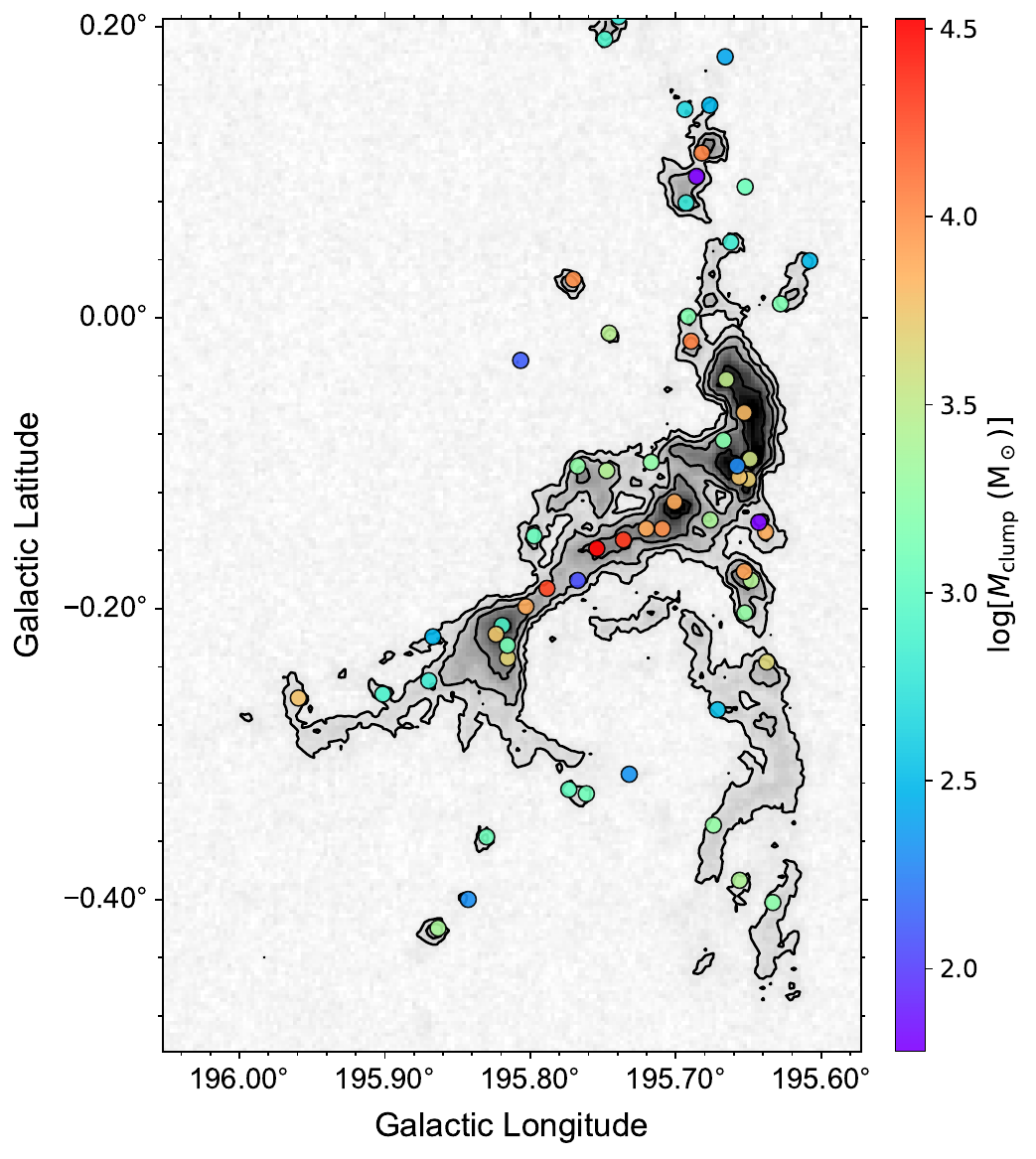}
    \end{minipage}
    \begin{minipage}[b]{0.45\textwidth}
        \centering
        \includegraphics[width=\linewidth]{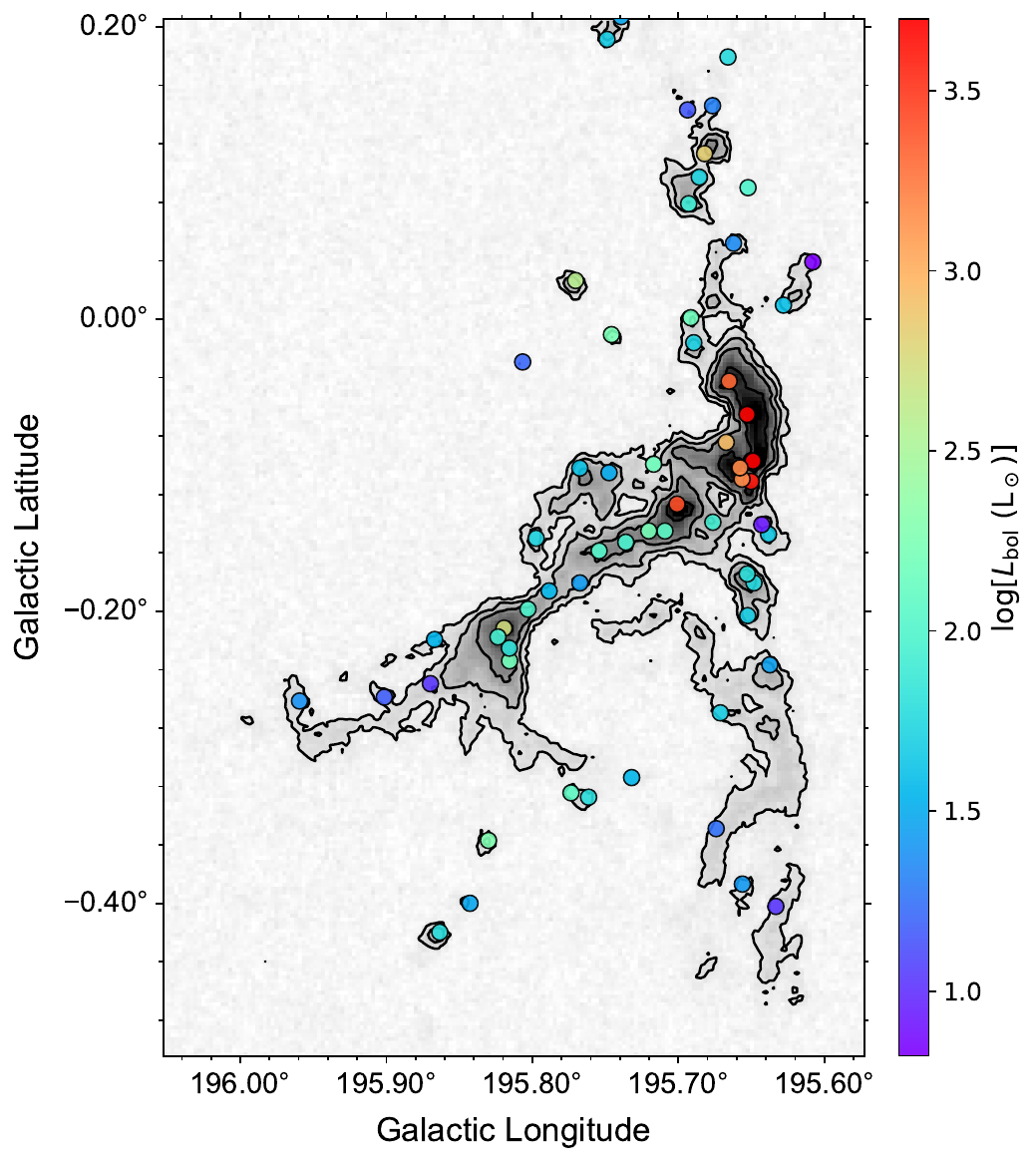}
    \end{minipage}

    \vspace{0.5em}

    \begin{minipage}[t]{0.45\textwidth}
        \centering
        \includegraphics[width=\linewidth]{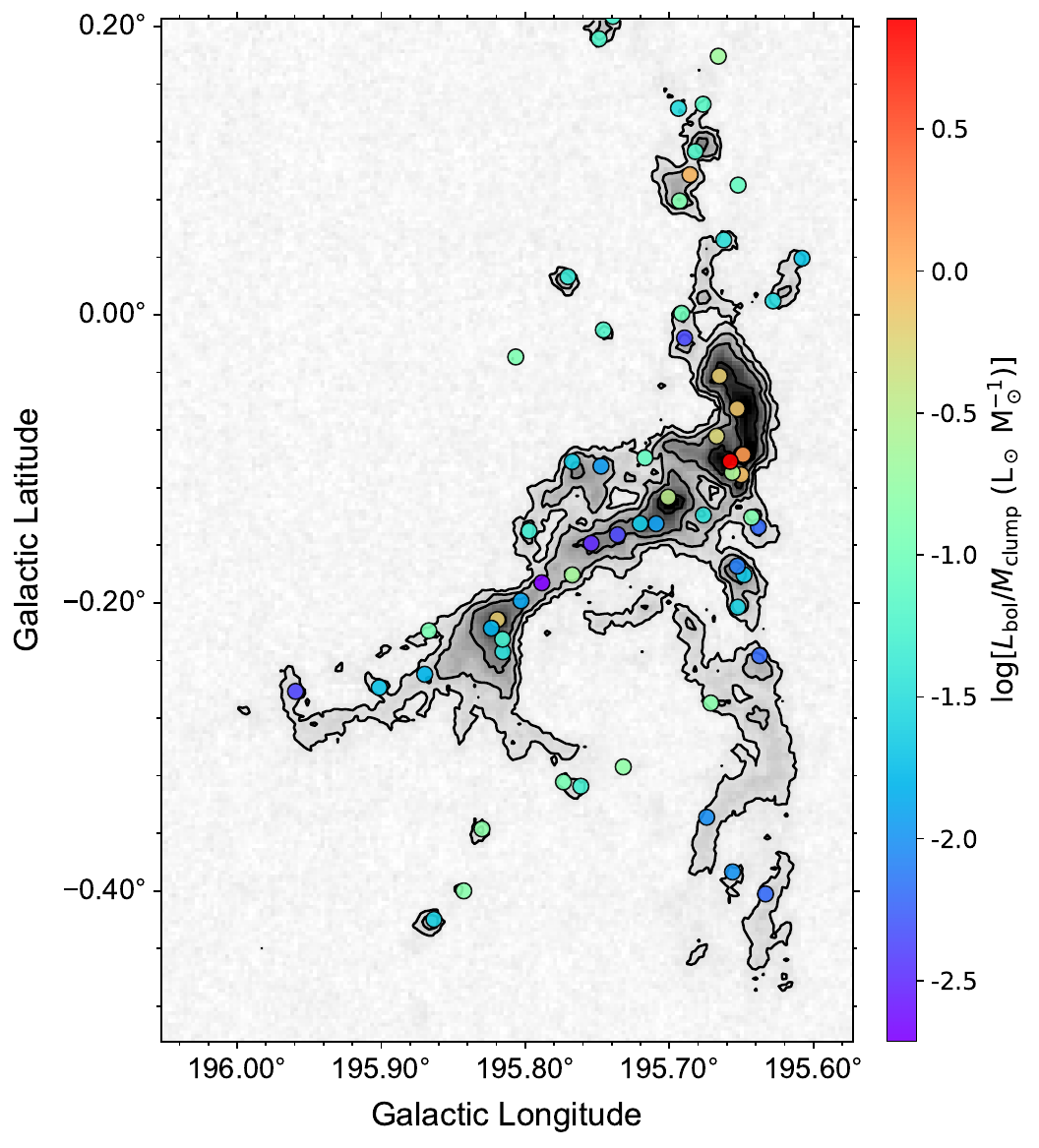}
    \end{minipage}
    \begin{minipage}{0.45\textwidth}
    \centering
    \begin{minipage}{0.80\textwidth}
     
     \vspace{-260pt}
     \hspace{30pt}
        \caption{Background map and contours as described for the upper panels of Fig.\,\ref{fig:clump_properties}. Filled circles again show the positions of the associated \higal clumps but in the upper left, upper right and lower left panels the colours indicate the clump mass, luminosity and $L_{\rm bol}/M_{\rm clump}$-ratio respectively. The concentration of the most massive clumps in the high column density region embedded in Filament\,1 and the most luminous clumps towards the central hub are evident; the radial behaviour of \lm for the protostellar population is examined quantitatively in Fig.\,\ref{fig:lm_ratio}.
     }
        \label{fig:three_panel}
        
    \end{minipage}
    \end{minipage}

\end{figure*}

\subsection{A giant hub-filament system in the outer Galactic disc}

The \complexnamelambda combines the morphology of a classical hub-filament system with the physical scale of a giant molecular filamentary complex. Its three principal filaments extend for approximately 34–42\,pc and converge on a dense central hub, while the full molecular complex spans approximately $35\times63$\,pc. The overall structure of the complex looks strikingly similar to SDC13 (\citealt{peretto2014}), although, it is a factor of ten larger in size, and is also substantially larger than the characteristic 10–20\,pc dimensions of most hub-filament systems identified by \citet{kumar2020}.

The complex is also unusual in its Galactic location. The adopted Gaia distance places it at \rgc{13.2}, making it one of the most distant well-resolved hub-filament systems from the Galactic centre. \citet{kumar2020} identified only a small number of HFS candidates outside the Solar circle, and these are generally located at much larger heliocentric distances. For example, HF09001 has a reported heliocentric distance of 23.2\,kpc \citep{kumar2020}, corresponding to a physical resolution of approximately 3.4\,pc for a 30-arcsec beam. The equivalent physical resolution for the \complexnamelambda is approximately 0.75\,pc.

The combination of its large physical extent, substantial molecular reservoir of $\sim1.6\times10^5$\,\msun, and comparatively high physical resolution provides an excellent opportunity to investigate the organisation of dense gas and star formation within a giant HFS in the outer Galactic disc.

\subsection{Star formation within the hub-filament system}
\label{sect:star_formation}

The spatial distributions of the protostellar and starless clump populations indicate that star formation is not uniformly distributed throughout the complex. Although gravitationally bound starless clumps are found along the full extent of the filamentary network, the protostellar population is preferentially concentrated towards the central hub (Fig.\,\ref{fig:surface_density}). The median projected distance from the hub is 6.65\,pc for the protostellar clumps and 15.83\,pc for the starless population, and a two-sided Mann-Whitney U test indicates that their radial distributions are significantly different (\pvalue{0.011}). The contrast is particularly pronounced within the central 10\,pc, where 10 of the 20 clumps are protostellar, corresponding to a protostellar fraction of 0.50, compared with only 6 of 41 (0.15) at larger offsets. The central concentration of protostellar clumps therefore appears to be a significant feature of the complex rather than simply a consequence of the particular 10-pc radius adopted for comparison.

The physical properties of the clumps show a similar radial dependence. Surface density, clump mass and bolometric luminosity all decrease significantly with increasing distance from the hub, with Spearman rank coefficients of $\rho_{\rm S}=-0.36$ ($p$-value = 0.0044), $-0.42$ (\pvalue{$6.8\times10^{-4}$}) and $-0.50$ (\pvalue{$4.1\times10^{-5}$}), respectively. The central region therefore preferentially contains the denser, more massive and more luminous clumps in the complex (Fig.\,\ref{fig:three_panel}). 

\begin{figure}
	\centering
        
        \includegraphics[width=0.45\textwidth, trim = 0cm 0cm 0cm 0cm, clip]{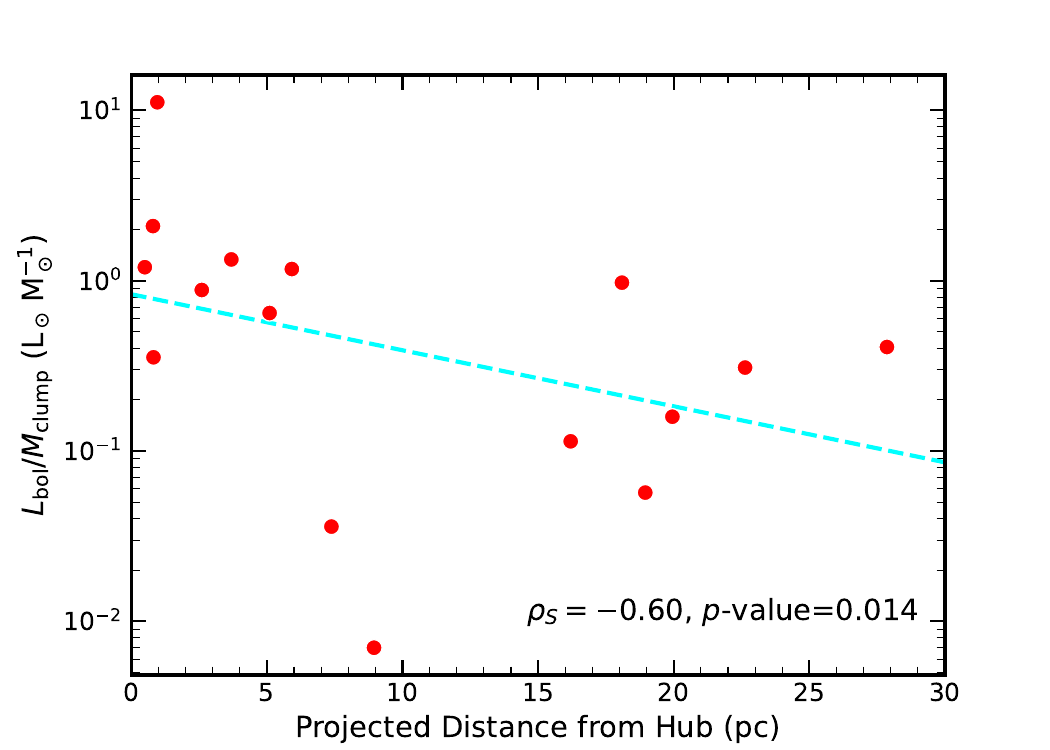}
       
        \caption{Bolometric luminosity-to-mass ratio of the 16 protostellar Hi-GAL clumps as a function of projected distance from the central hub. The dashed line shows a linear fit to the logarithm of \lbol/\mclump and is included to illustrate the overall trend. The Spearman rank correlation coefficient and corresponding $p$-value are given in the lower-right corner.}
		\label{fig:lm_ratio}
	
\end{figure}

There is also evidence that the protostellar population itself exhibits an evolutionary gradient. As expected, the protostellar clumps have substantially higher luminosity-to-mass ratios than the starless population, with median $L_{\rm bol}/M_{\rm clump}$ values of 0.53 and 0.057\,L$_\odot$\,M$_\odot^{-1}$, respectively. \lm provides a commonly used, distance-independent indicator of clump evolution, generally increasing as the luminosity of the embedded stellar population rises relative to the mass of its natal clump (\citealt{molinari2008,urquhart2014_atlas}). Considering only the 16 protostellar clumps, \lm decreases significantly with increasing projected distance from the hub ($\rho_{\rm S}=-0.60$, \pvalue{0.014}; Fig.\,\ref{fig:lm_ratio}), indicating that the most evolutionarily advanced protostellar clumps are preferentially located towards the central region. A similar, although weaker, trend is seen in dust temperature ($\rho_{\rm S}=-0.47$, \pvalue{0.064}). Importantly, the \lm-offset relation becomes stronger after controlling for clump mass, with a partial Spearman rank coefficient of $\rho_{\rm partial}=-0.84$. A $10^5$-iteration permutation test gives \pvalue{$1.6\times10^{-4}$}, demonstrating that the observed evolutionary gradient is unlikely to result solely from the preferential concentration of more massive clumps towards the hub. Indeed, many of the most massive clumps are located along the length of Filament\,1 away from the central hub.

The concentration of evolved sources towards the hub is particularly significant in the context of high-mass star formation. The central region already contains an embedded stellar cluster, two luminous protostellar clumps associated with candidate high-mass protostellar objects ($L_{\rm bol}>10^3$\,\lsun), and the \hii region G195.648$-$00.106, whose radio continuum emission is consistent with ionisation by a $\sim15$\,\msun ZAMS star (Sect.\,\ref{sect:g195_complex}). High-mass star formation is therefore already established within the hub. At the same time, 18 of the 61 associated clumps have masses exceeding $10^3$\,\msun, demonstrating that a substantial reservoir of massive dense material remains available for further star formation. As previously mentioned, several of the most massive clumps are concentrated within the high-column-density region along Filament\,1 and have relatively low \lm ratios, indicating that they are at an earlier evolutionary stage (see upper and lower left panel of Fig.\,\ref{fig:three_panel}). Thus, while the central hub currently hosts the most evolved and active star formation, the  young massive clumps distributed elsewhere in the filamentary network provide considerable potential for future high-mass star formation.

Taken together, these results reveal a clear spatial organisation in both the incidence and evolutionary state of star formation across the \complexnamelambda. The central hub is the current focus of the most advanced star formation, while the surrounding filamentary network contains a substantial reservoir of gravitationally bound and massive clumps at earlier evolutionary stages. The observations do not, however, establish whether this evolutionary segregation results from the dynamical transport of material towards the hub, or whether star formation commenced earlier or proceeds more rapidly there as a consequence of the higher gas and clump densities. Whether the observed spatial and evolutionary organisation is related to material transport through the filamentary network is considered in the following subsection.

\subsection{Large-scale gas motions}
\label{sect:large_scale_motions}

The clump velocities and the $^{12}$CO velocity field reveal coherent large-scale kinematic structure across the complex. In general, the velocity gradients appear steepest near the central hub and become shallower with increasing distance along the filaments. Similar velocity gradients along filaments have been interpreted in other HFSs as evidence for longitudinal gas flows towards central hubs (e.g. \citealt{peretto2014,williams2018}).  In particular, \citet{peretto2014} found that the gas velocity anti-correlates with velocity dispersion along the SDC13 filaments -- the velocity gradients are smaller at the filament ends and increase towards the hub centre -- which they attributed to gravitational acceleration along the filaments. The velocity dispersion maximum at the hub centre is naturally explained by the collision of multiple converging flows (\citealt{williams2018}). Our \co{2}{1} data show a qualitatively similar velocity pattern, although we caution that the optically thick nature of \co{2}{1} may introduce additional complexities.

The position-velocity diagrams presented in Fig.\,\ref{fig:pv_diagrams} provide some support for organised motions along the filaments, but the interpretation is not unambiguous. The fitted gradients differ between the three structures, and the molecular emission exhibits substantial local velocity substructure. In particular, the velocity behaviour along individual filaments is not always well represented by a single linear gradient. The measured line-of-sight gradients are also sensitive to the unknown inclinations of the filaments relative to the plane of the sky.

The observed kinematics are therefore consistent with organised large-scale motions associated with the hub-filament geometry, including the possibility of longitudinal flows towards the hub, but the current observations do not uniquely establish the direction or physical origin of these motions. Alternative contributions from global cloud contraction, local gravitational collapse, projection effects and multiple velocity components cannot presently be excluded.

The coexistence of coherent large-scale velocity structure and smaller-scale kinematic variations associated with individual condensations is qualitatively consistent with hierarchical models of molecular-cloud evolution, in which global contraction and local collapse can occur simultaneously (e.g. \citealt{Gomez2014,Vazquez2009,Vazquez2019,Vazquez2025,Camacho2020}). Higher spatial and spectral resolution observations using suitable dense-gas tracers will be required to distinguish between these possibilities and to determine whether material is being transported systematically along the filaments towards the central hub.

\subsection{Implications for star formation in the outer Galaxy}

One of the principal motivations for studying the \complexnamelambda is its location at a Galactocentric radius of 13.2\,kpc. Most well-studied hub-filament systems are located considerably closer to the Galactic centre, while the outer Galactic disc is characterised by environmental conditions that differ from those prevailing in the inner Milky Way.

Despite its outer-Galaxy location, this complex exhibits many of the principal characteristics associated with hub-filament systems elsewhere in the Galaxy: multiple elongated molecular structures converge on a high-column-density hub; dense clumps are distributed throughout the filament network; protostellar objects are preferentially concentrated towards the hub; and the molecular gas exhibits coherent velocity structure on scales of several tens of parsecs.

These similarities suggest that the organisation of molecular gas into hub-filament structures and the preferential concentration of star formation towards their intersections can operate efficiently at large Galactocentric radii. The \complexnamelambda therefore provides an important benchmark for investigating whether the physical processes responsible for clustered and high-mass star formation depend strongly on Galactic environment. 

The established high-mass star formation within this complex is particularly notable given the declining incidence of high-mass star-forming regions towards the outer Galactic disc \citep{urquhart2014_rms,djordjevic2019}. This demonstrates that the outer-Galaxy environment is capable not only of producing a giant, highly organised hub-filament system, but also of assembling sufficient dense material to support sustained high-mass star formation (as discussed in Sect.\,\ref{sect:star_formation}).

Determining whether the \complexnamelambda is representative of star-forming structures at large Galactocentric radii requires a systematic analysis of the wider OGHReS complex population, which is beyond the scope of the present study.

\section{Conclusions}
\label{sect:conclusions}

We have presented the discovery and characterisation of the \complexnamelambda, a giant hub-filament system identified in the OGHReS survey. This complex occupies an unusual region of parameter space. It combines the size of giant filamentary complexes with the morphology of a classical hub-filament system, while lying in one of the most remote star-forming environments yet studied at high spatial resolution. As such it provides an unprecedented laboratory for testing theories of clustered star formation under the lower-metallicity conditions characteristic of the outer Milky Way. Our main conclusions are as follows.

\begin{enumerate}

\item The \complexnamelambda is a spatially and kinematically coherent molecular complex containing 61 \higal clumps with a mean velocity of $V_{\rm LSR}=32.85$\,\kms and a velocity dispersion of 1.97\,\kms. Adopting the $Gaia$ distance of 5.2$^{+0.8}_{-0.6}$\,kpc to MWC 137, the complex lies at a Galactocentric radius of 13.2\,kpc and spans approximately $35\times63$\,pc, making it one of the largest well-resolved hub-filament systems identified in the outer Galactic disc. Furthermore, with a metallicity of $\approx0.4\,Z_\odot$ this complex opens a window to studying HFS and star formation in low-metallicity environments and provides a link to extragalactic star formation studies (e.g. \citealt{Thilker2023}). \\

\item The molecular emission is organised into three prominent filaments that converge on a dense central hub. The total molecular gas mass is estimated to be $\sim1.6\times10^{5}$\,\msun, while approximately $5\times10^{4}$\,\msun is contained in dense \higal clumps, corresponding to a dense-gas fraction of $\sim 0.3$. The individual filaments have lengths of 34-42\,pc and widths of 2.1-3.1\,pc, giving aspect ratios of 13-20.\\

\item The protostellar population is significantly more concentrated towards the hub than the starless population (median offsets 6.65 versus 15.83\,pc; \pvalue{0.011}). Among the protostellar population, \lm also decreases significantly with hub offset, and this relation remains significant after controlling for clump mass ($\rho_{\rm partial} = -0.84$, permutation \pvalue{$1.6\times10^{-4}$}).\\

\item The \complexnamelambda is already actively forming high-mass stars while retaining considerable potential for further high-mass star formation. The central hub contains an embedded stellar cluster, two protostellar clumps with $L_{\rm bol}>10^3$\,\lsun, and an \hii region whose radio continuum emission is consistent with ionisation by a $\sim15$\,\msun ZAMS star. Moreover, 18 of the 61 associated clumps have masses exceeding $10^3$\,\msun, demonstrating that a substantial reservoir of massive dense structures remains distributed throughout the complex. This complex therefore represents not only a giant hub-filament system, but an active high-mass star-forming complex with the potential to sustain further massive-star formation.\\

\item The molecular gas exhibits coherent large-scale velocity structure, with velocities generally increasing away from the hub along two of the three principal filaments. These motions are consistent with organised gas flows within the hub-filament network but do not uniquely demonstrate longitudinal accretion. Taken together with the evolutionary segregation of the dense-clump population, the \complexnamelambda provides a particularly well-resolved example of hierarchical and high-mass star formation operating at large Galactocentric radius.

\end{enumerate}

\section*{Acknowledgements}

We thank the referee for their comments and suggestions that have helped clarify a number of aspects of this work.  AYY acknowledges the support from the National SKA Program of China (2025SKA0140100), and National Natural Science Foundation of China (NSFC) grants No.\,12303031. MAT acknowledges support from STFC grant awards ST/R000905/1 and ST/W00125X/1. DC acknowledges support from the Deutsche Forschungsgemeinschaft (DFG; German Research Foundation) through Collaborative Research Centre SFB 1601 (subproject B3; grant No.\,500700252) and Germany’s Excellence Strategy, EXC 3037 ``Our Dynamic Universe'' (grant No.\,533607693). AK and MF acknowledge support from the Polish National Science Center SONATA BIS grant No.\,2024/54/E/ST9/00314 and SONATA grant No.\,2022/47/D/ST9/00419 respectively.

\section*{Data Availability}

The \higal\ catalogue used for this work can be found at \href{http://vialactea.iaps.inaf.it/vialactea/public/}{http://vialactea.iaps.inaf.it/vialactea/public/} and the OGHReS catalogue used here can be found on CDS.


\bibliographystyle{mnras}
\bibliography{urquhart2025}

\end{document}